\documentclass[reviewcopy]{elsart}
\usepackage[]{fontenc}
\usepackage[latin1]{inputenc}
\usepackage{floatflt}
\usepackage{graphicx}
\makeatletter
\newcommand{\noun}[1]{\textsc{#1}}

\makeatother
\begin{document}
\begin{frontmatter}
\title{Cosmological gamma ray and neutrino backgrounds due to neutralino dark matter annihilation}
\author{Dominik Elsässer\corauthref{cor}}
\corauth[cor]{Corresponding Author}
\ead{elsaesser@astro.uni-wuerzburg.de}
 and 
\author{Karl Mannheim}
\ead{mannheim@astro.uni-wuerzburg.de}
\address{Institut für Theoretische Physik und Astrophysik, Universität Würzburg, Am Hubland, 97074 Würzburg, Deutschland }

\date{Accepted Version; 23.04.2004}

\begin{abstract}
We compute the cosmological background radiation of gamma rays and neutrinos
due to neutralino annihilation in evolving dark matter halos, assuming the
observed dark matter is comprised of thermally excited neutralinos in the
MSSM. The spectrum of this gamma-ray background radiation does not show
strong annihilation line features, but could amount to a significant fraction
of the extragalactic gamma ray continuum flux observed by EGRET above a few GeV.
The corresponding cosmological neutrino background is weak compared to the
atmospheric foreground. Assuming full mixing, however, the cosmological
tau-neutrino background could be detectable with a flavor-discriminating
neutrino telescope in the energy range 10 GeV -- 1 TeV. A small anisotropy of
the background radiation is expected, reflecting the local clustering of dark
matter halos along the supergalactic plane.
\end{abstract}


\end{frontmatter}

\section{Introduction}
The nature of the dark matter in the Universe is one of the biggest
mysteries in modern astrophysics \cite{dm-primack}. Of the numerous proposed constituents,
the supersymmetric WIMP (commonly believed to be the lightest neutralino $\mathrm{\chi_{1}^{0}}$) is perhaps the most convincing candidate \cite{key-3}. A possible
way to find signatures of neutralino dark matter is to look for gamma
rays being produced due to the self-annihilation of the neutralinos
\cite{key-31}. Detection of a gamma ray line signal at energies above a few GeV would constitute
an immediate proof of new physics beyond the Standard Model. However, the process $\mathrm{\chi\chi\rightarrow\gamma\gamma}$ is loop-suppressed \cite{key-loop},
and so these intensities are expected to be very small. Therefore,
it has been proposed to search for continuum gamma rays resulting from the production and following decay of $\mathrm{\pi}$-mesons (and to a lesser extent also K-mesons) in the annihilation process. Searches
for continuum gamma rays originating from the dark halo of our
own galaxy, but also from several extragalactic targets (including dSph-galaxies as well as M87, M31 and other nearby galaxies)
have been discussed (see e.g. \cite{Baltz,key-32,key-33,dixon} for a review). 
Since neutralino annihilation has taken place throughout the Universe
essentially since the freeze-out of the dark matter, a homogeneous
and - at a first glance - isotropic background of annihilation products should also exist \cite{schramm,BEU,key-21}. Comparing the measured extragalactic gamma ray background and predictions from blazar models \cite{chiang,salamon}, one finds that there might be room for additional contributions.
In this paper, we focus on the continuum gamma ray background (Section 2), as
well as on expected high-energy neutrino backgrounds (Section 3). These annihilation induced backgrounds are significant since they are the only cases of extragalactic backgrounds scaling with the matter density squared, instead of just the matter density, as for all conventional backgrounds. We discuss a possible anisotropy on large angular scales in those backgrounds owing to the anisotropic mass distribution in the local Universe, which is a potentially distinct signature of neutralino annihilation. Furthermore, in the case of supersymmetric dark matter, these intensities are irreducible backgrounds for all observations, including targeted SUSY-searches with next-generation IACTs (Imaging Air shower Cerenkov Telescopes), which are operated in ON - OFF detection mode for background suppression.  

\section{Gamma rays due to neutralino annihilations}

The differential gamma ray intensity per solid angle due to
neutralino annihilations in any given dark matter distribution is 
\begin{equation}
\mathrm{\Phi_{\gamma}(\varepsilon)=\frac{1}{4\pi}\times\frac{\frac{1}{2}\left\langle\sigma v\right\rangle }{m_{\chi}^{2}}\times\int\rho^{2}_{\chi}\times\kappa\left[\varepsilon\left(1+z\right),z\right] \times df\,[\varepsilon(1+z)]\;\frac{c\,dt}{dz}\;dz}\label{eq:1}\;\;,\end{equation}

where $\mathrm{df\,[\varepsilon(1+z)]}$ is the differential energy distribution of
the produced \linebreak gamma photons per annihilation event, 
$\mathrm{\left\langle \sigma v\right\rangle }$
is the self-annihilation cross section averaged over the thermal distribution of
the dark matter particles, and $\mathrm{m_{\chi}}$ is the particle mass. $\mathrm{\kappa\left[\varepsilon\left(1+z\right),z\right]}$
parameterizes the gamma ray attenuation on cosmological scales, and $\mathrm{\int}$$\rho^{2}_{\chi}\,\mathrm{ds}$ with $\mathrm{ds=\frac{c\,dt}{dz}\;dz}$
is the integral along the line of sight over the dark matter density
squared. The factor $\mathrm{1/2}\;$ is included to compensate for evaluating
twice each initial state going into ${\mathrm{\left\langle \sigma v\right\rangle}}$
\cite{key-21}. While $\mathrm{{\frac{1}{2}\left\langle \sigma v\right\rangle }/{m_{\chi}²}}$ and the differential energy distribution are fixed by the choice of the SUSY-model, the integral over the dark matter density depends on the astrophysical abundance and distribution of the dark matter, specifically the value of $\mathrm{\Omega_{DM}}$ and the
dominant type of dark matter halo and subhalo profile. $\mathrm{{\varepsilon}=E/(z+1)}$ is the observed energy of a photon emitted at energy E. To calculate
the total cosmological intensity from dark matter annihilations, we start with
the finite number of neutralinos produced during thermal freeze-out, which are
depleted thereafter following the Boltzmann equation $\;\mathrm{dn_{\chi}/dt=-\left\langle \sigma v\right\rangle\;n_{\chi}^{2}\,(1+z)^{3}}$. \,This approach has also been employed in \cite{BEU} and yields 

\begin{eqnarray}
\mathrm{\Phi_{\gamma}^{Neutralino}\,(\varepsilon)=\frac{c}{4\pi H_{0}}\times\frac{\frac{1}{2}\left\langle \sigma v\right\rangle \;\Omega²_{DM}\;\rho²_{crit}}{m_{\chi}²}}\times\nonumber\\
\mathrm{\int_{0}^{z_{max}}dz\frac{\left(1+z\right)^{3}\times\kappa\left[\varepsilon\left(1+z\right),z\right]\times\Gamma\left(z\right)}{h\left(z\right)}
\times df\,(\varepsilon(1+z))}\label{eq:2} .\end{eqnarray}

In existing literature, most attention has been devoted to the cosmological annihilation into gamma ray lines, while we will take a different approach and focus on the characteristics and potentially detectable features of the extragalactic continuum signal. Therefore, some adaptations of the ingredients in Eq. \ref{eq:2} have to be included. On the "astrophysical side", $\mathrm{\Gamma\left(z\right)}$ denotes the intensity multiplier function obtained from simulations of structure formation \cite{key-5,ens}. $\mathrm{h\left(z\right)}$ is
given through 
\begin{equation}
\mathrm{h\left(z\right)=\sqrt{\Omega_{M}\left(1+z\right)^{3}+\Omega_{K}\left(1+z\right)^{2}+\Omega_{\Lambda}}}\;\;,
\end{equation}
and for the "concordance" cosmology we will assume in the following
that $\mathrm{\Omega_{\chi}\cong\Omega_{DM}\cong0.23}$, $\mathrm{\Omega_{M}\cong0.27}$,
$\mathrm{\Omega_{K}=0}$ and $\mathrm{\Omega_{\Lambda}\cong0.73}$ \cite{key-13}. For the Hubble-Parameter $\mathrm{H_{0}}$ we will work with the current best fit value of\\ 
$\mathrm{H_{0}=71\;{km}\;{s }^{-1}\;{Mpc}^{-1}}$
\cite{key-13}. $\rho_{crit}\;$ is$\;$ $\mathrm{3H^{2}/8\pi G\;}$, the overall (critical) density of the Universe for $\;\Omega_{0}=1$. As has been pointed out e.g. in \cite{key-4}, the annihilation induced intensity heavily depends on the amount of structure in
the dark matter. Instead of a constant "clumpiness factor", in this paper we include physics-motivated calculations of the z-dependence of the intensity multiplication function. For redshifts up to $\mathrm{z = 20}$ for $\mathrm{\Gamma\left(z\right)}$ we will use calculations of the "dimensionless flux multiplier" from \cite{key-5}, which trace the process of structure and substructure formation. The results are in part based on the calculations and computer code presented in \cite{ens}. At higher redshifts, knowledge about structure in the Universe is rather limited, and we conservatively assume a completely smooth and structureless dark matter distribution at $\mathrm{z > 20}$. One of the remaining questions is how much of the already formed halos are destroyed due to processes like tidal stripping and merger events. If this is the case for a sufficient number of halos, the relative intensity contribution due to annihilation events might actually peak quite early and then decline again, as indicated in \cite{key-5}. However, as the authors acknowledge therein, this effect might well be overestimated in \cite{key-5}, e. g. because in N-body simulations the central regions of merging halos tend to survive. A more sophisticated treatment of the details of structure formation history should definitely be rewarding and perhaps settle this question. It should be noted that if this peak is in fact less prominent than assumed here, the intensity from the relatively local Universe will be even more significant with respect to the overall annihilation induced intensity. A lower mass cutoff of $\mathrm{10^{6}}$ solar masses for the dark halos is assumed, which represents the mass scale down to which the paradigm of hierarchical structure formation presently seems to be well established. We study two representative cases of dark matter halo profiles, the Moore et al. profile \cite{key-6}
\begin{equation}
\mathrm{\rho\left(r\right)=\frac{\rho_{0}}{\left(\frac{r}{r_{scale}}\right)^{1.5}\left[1+\left(\frac{r}{r_{scale}}\right)^{1.5}\right]}}\;\;\label{eq:3a}\end{equation}
and the NFW-type halo profile \cite{key-34}
\begin{equation}
\mathrm{\rho\left(r\right)=\frac{\rho_{0}}{\left(\frac{r}{r_{scale}}\right)\left(1+\frac{r}{r_{scale}}\right)^{2}}}\;\;.\label{eq:3b}\end{equation}

In both cases, the Sheth and Tormen (ST) mass function presented in \cite{st} is assumed. 
We notice that in the recent literature, while some cases of (elliptical)
galaxies with very shallow and low-mass dark halos have been reported
\cite{key-7}, there has been convincing evidence of the
validity of $\mathrm{r^{-\beta}}$-like dark matter halo profiles from x-ray
observations of Abell galaxy clusters, with $\beta = 1.35 \pm 0.21$ in the case of A2589 \cite{key-8,key-9}. At least in the case of this cluster, the observed profile is not quite as steep as the Moore et al. profile, but has some excess mass with respect to the pure NFW case. As has also been suggested in \cite{key-5}, the Moore et al. case can therefore be considered an optimistic but not extreme scenario. For none of our neutralino models this scenario is constrained by expected gamma ray intensities from substructure in the Milky Way halo, if the external radius of the Milky Way clumps is taken to be their virial radius \cite{olinto2,key-21}. Expected extragalactic radio intensities from annihilation products of neutralinos in our models also are not larger than the intensities due to normal and radio galaxies \cite{radio}. Compared to the enhancement factors employed in previous literature (e.g. \cite{BEU}), the calculations used here result in a nearly an order of magnitude larger  present-day intensity enhancement of $\mathrm{1.5\times10^{7}}$ for the Moore et al. halo profile.

For the cosmological distances of interest here, high
energy gamma rays cannot propagate freely through the Universe owing to pair production with infrared-to-ultraviolet background photons. The
details of gamma ray attenuation are very important for observations
at cosmological distances, and have been subject to scientific debate
in recent years (e.g. \cite{mrf-stecker,mrf-primack,key-10}).
The optical depth due to pair creation in the metagalactic radiation
field (MRF) for a source at redshift $\mathrm{z_{q}}$ and at the observed energy
$\mathrm{\varepsilon}$ is calculated to be

\begin{equation}
\mathrm{\tau_{\gamma\gamma}({\varepsilon},z_{q})=c\int_{0}^{z_{q}}\int_{0}^{2}\int_{0}^{\infty}\frac{dl}{dz^{'}}\frac{\mu}{2}\times n(z,E)\times\sigma_{\gamma\gamma}({\varepsilon},E,\mu,z^{'})\;dE\;d\mu \;dz^{'}},\label{eq:4}\end{equation}
where $\mathrm{{dl}/{dz^{'}}}$ is the cosmological line element, 
$\mathrm{\theta}$ is the
angle between interacting photons, $\mathrm{\mu=cos\left(\theta\right)}$,
n(z,$\mathrm{E}$) is the MRF photon number density and $\mathrm{\sigma_{\gamma\gamma}\left(\varepsilon,E,\mu,z\right)}$ the
pair production cross section. The Fazio-Stecker-Relation presented in \cite{key-10}
allows us to estimate the redshift
 $\mathrm{z_{cutoff}(\left[\varepsilon\left(1+z\right)\right],z)}$ 
at which
the optical depth for a gamma photon of a given energy will be of
the order one, so that we get \begin{equation}
\mathrm{\kappa\left[\varepsilon\left(1+z\right),z\right]=exp\left[{-\frac{z}{z_{cutoff}[\varepsilon(1+z),z]}}\right]}\label{eq:5}\end{equation}
for the absorption of the neutralino-induced gamma rays. We use the best-fit-model Fazio Stecker Relation from \cite{key-10}. Since reemission
of photons with initial energies of some ten GeV occurs below
100 MeV, reemission will be neglected for our analysis of the high-energy gamma
ray part of the spectrum. The high energy gamma rays are potentially most interesting because in this regime, the SUSY-contribution might have the highest significance against the steep conventional backgrounds.
Since the calculations from \cite{key-10} depend on the properties of the metagalactic radiation field, and thus on the star formation history, they cannot reliably be extended beyond a redshift of roughly five. For redshifts larger than five and up to $\mathrm{z = 20}$, which for the gamma ray background we take as the upper limit $\mathrm{z_{max}}$ of the integration in Eq. (\ref{eq:2}), we use more robust estimates for the gamma absorption at high redshifts from \cite{zdz}. Due to the combined effect of redshift and structure formation, our end result is not really sensitive to the details of gamma ray attenuation at high redshifts.
Now, having discussed the astrophysical constituents in Eq. (\ref{eq:2}),
we turn to the SUSY-parameter set. Even in the Minimal Supersymmetric
Standard Model (MSSM), there are at least seven free parameters relevant to dark matter studies. We use the \noun{DarkSusy} numerical package \cite{key-16}
to make large scans over the MSSM parameter space. The models are
required to produce a neutralino not excluded by current experimental
results \cite{key-17}. Furthermore,
the considered models are required to produce $\mathrm{0.025\leq\Omega_{\chi}h^{2}\leq0.176}$.
The higher limit is a little larger than the present best-fit value for $\mathrm{\Omega_{DM}}$ \cite{key-13} to include possible residual
uncertainties in the concordance model. The lower limit accounts
for the fact that the dark matter in principle need not be constituted
entirely of neutralinos, but they should make at least a significant
contribution. For models that will not produce $\mathrm{\Omega_{\chi}\cong\Omega_{DM}}$ we rescale the resulting intensities accordingly.
The differential energy
distribution for each model can also be calculated using \noun{DarkSusy}.
Fig. 1 shows a scatter plot of more than 3000 neutralino models that
match the criteria mentioned above. For each model we plot
the neutralino rest mass and the parameter "R", which we define
as the ratio of the continuum gamma ray intensity created by the given
model assuming the Moore et al. halo profile, and the extragalactic EGRET intensity ($\mathrm{9\times10^{-10}\;cm^{-2}\;s^{-1}\;sr^{-1}\;GeV^{-1}}$) \cite{key-EGRET}, at an energy of 30 GeV: \begin{equation}
\mathrm{R=\frac{\Phi_{\gamma}^{Neutralino}}{\Phi_{total,EGRET}}\mid_{30GeV}}\label{eq:6} .\end{equation}

At this energy, the contributions from highly productive
models in the Moore et al. case will reach the ten-percent level, indicating that there might well be a noticeable component due to WIMP annihilation in the measured EGRB.

\begin{figure}[!t]
	\begin{center}
		\includegraphics[width=12cm,height=8.4cm]{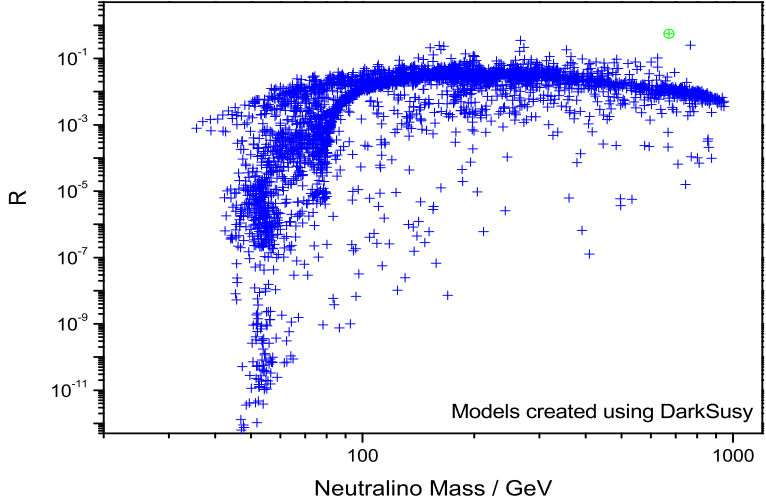}
  \end{center}
	\caption{R as defined in Section 2 for $\mathrm{3\times10^{3}}$ cosmologically valid models plotted over neutralino rest mass. The Moore et al. halo profile case is assumed. The circle denotes the highly productive model discussed in detail in Section 2.}
	\label{fig:Fig1}
\end{figure}

\begin{figure}[hbp]
	\begin{center}
		\includegraphics[width=12cm,height=8.4cm]{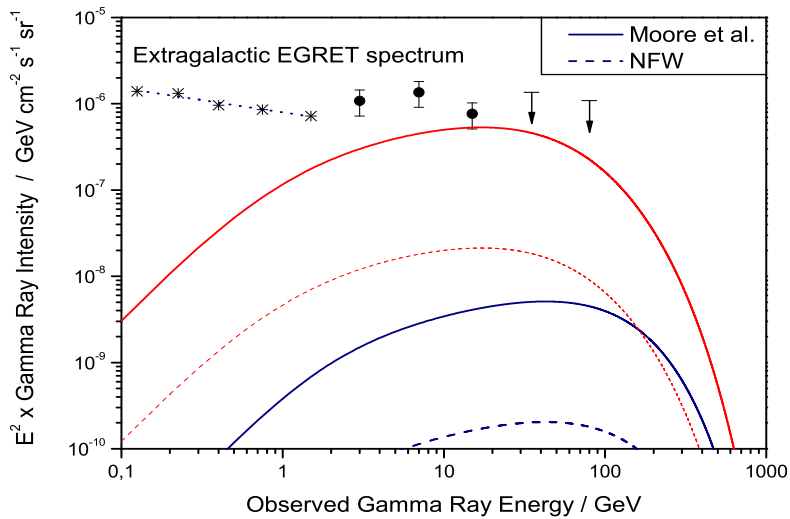}
	\end{center}
	\caption{Continuum gamma ray intensities in the Moore et al. / NFW case over observed energy for the neutralino model described in Section 2; the upper solid / dashed lines correspond to the total SUSY-induced intensity in the respective case, and the lower solid / dashed lines represent the estimated contribution from the local Universe (up to z = 0.01). The extragalactic EGRET intensity from \cite{key-reimer,key-EGRET}(stars / dots) is also shown for comparison. }
	\label{fig:Fig2}
\end{figure}

In Fig. 2, for one model with $\mathrm{m_{\chi}=672\;GeV}$ and $\mathrm{\left\langle \sigma v\right\rangle =6.1\times10^{-25}\;{cm^{3}}\;{s}^{-1}}$ we plot the cosmological gamma ray
signal from neutralino annihilations as a function of observed energy.
For the measured extragalactic EGRET  background, we use the new determination by Strong et al. \cite{key-reimer}. The bulk of the EGRB is likely due to unresolved, faint gamma ray emitting AGN (blazars) \cite{chiang,salamon}.  However, as Fig. 2 indicates, a dominant cosmological component at high energies could also be produced by cuspy halo profiles, a high self-annihilation cross section and a neutralino mass in the range of (500--900)GeV. 
As can be concluded from Fig. 1, for many of the calculated parameter sets and especially in the NFW case of halo profile, the neutralino induced intensity will however not be the dominating component in the extragalactic gamma ray background. Thus, for singling out this signal from the conventional background, highly selective criteria will have to be employed. Such a criterion can be a significant branching ratio into the gamma ray line channels. Almost in all cases however this line intensity
will even be much smaller than the continuum intensities presented above,
and identifying an asymmetrically broadened and very weak
line signal at an unknown energy in many cases might not be possible. Still, there
can be additional characteristics to the neutralino annihilation
induced signal. In Fig. 2 the upper solid / dashed lines show the total neutralino annihilation induced background in our two cases of halo profile, and the lower solid / dashed lines correspond to the estimated contribution from the "local" (up to z = 0.01) Universe. This is done to point out that in case of a noticeable anisotropic dark
matter distribution in the local Universe, this anisotropy will lead
to an anisotropic extragalactic continuum intensity. From \cite{key-29} we extract the spatial characteristics of the local Universe, and the locally produced gamma ray intensity from the direction of the "supergalactic plane" is rescaled with the overdensity $\mathrm{{\delta}_{40}=0.48}$ for a supergalactic scale radius of 40 Mpc. Mainly
due to redshift, this "foreground" becomes more distinguished
at higher energies, and for the highly productive model presented
in Fig. 2 in the Moore et al. case it will reach approximately percent level respective to the total EGRB at 30 GeV (assuming the measured total extragalactic EGRET intensity of $\mathrm{9\times10^{-10}\;cm^{-2}\;s^{-1}\;sr^{-1}\;GeV^{-1}}$ at 30 GeV). Since the bulk of the measured EGRB is presently assumed to be due to more distant blazar sources, the anisotropy tracing the supergalactic plane might be a distinct feature of the intensity component produced by dark matter annihilations. In the NFW case the absolute intensities would be more than an order of magnitude lower, but the anisotropic behavior of the annihilation intensity itself is not sensitive to the details of the assumed profile. Such an anisotropy could in principle be detectable for next-generation satellite experiments (e.g. GLAST).
Owing to the strong $\rho^{2}$-scaling of the annihilation intensity, one would actually expect to "resolve" a significant part of the mass concentrations in the supergalactic plane, so that the discussed anisotropy should in fact be even more prominent than our calculations indicate. 
To study the effects of gamma ray absorption, in Fig. 3 we plot the calculated extragalactic intensities from three fiducial neutralino models with masses ranging from 100 GeV to 5 TeV. For each model we plot the relatively nearby (up to z = 0.03) and the "cosmological" (z larger than 0.03) contributions in comparison. The intensity becomes increasingly suppressed with rising particle mass because of declining number density. For heavier neutralinos, gamma ray absorption becomes the limiting factor for the observed intensity, and for gamma energies above the 1000 GeV range, the intensity from the relatively nearby Universe actually exceeds the cosmological contribution. For annihilating WIMPs with masses of 1 TeV and beyond, this effect would make the proposed anisotropy even more significant. At energies beyond the 1TeV-scale, the conventional extragalactic background from far-away sources is also expected to vanish due to gamma ray attenuation. 

\begin{figure}
	\begin{center}
		\includegraphics[width=12cm,height=8.4cm]{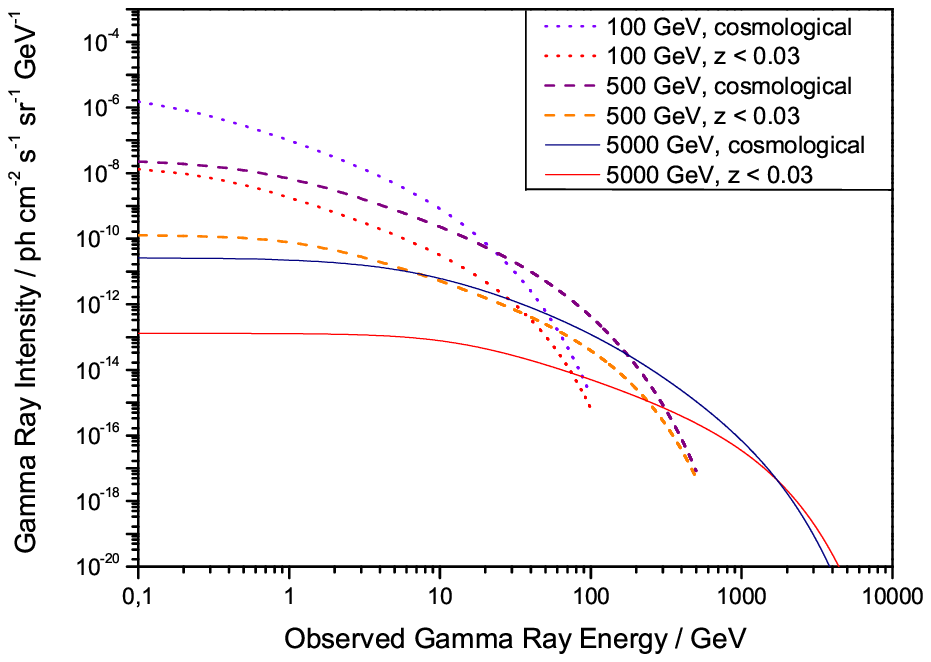}
	\end{center}
	\caption{Calculated extragalactic gamma ray intensities for three fiducial neutralino models with $\mathrm{\left\langle \sigma v\right\rangle =1\times10^{-26}\;{cm^{3}}\;{s}^{-1}}$} and masses of $\mathrm{m_{\chi}=100\;GeV,\;500\,GeV\;and\;5\,TeV}$. The Moore et al. case of halo profile is assumed.
	\label{fig:Fig3}
\end{figure}

\begin{figure}[!ht]
	\begin{center}
		\includegraphics[width=12cm,height=8.4cm]{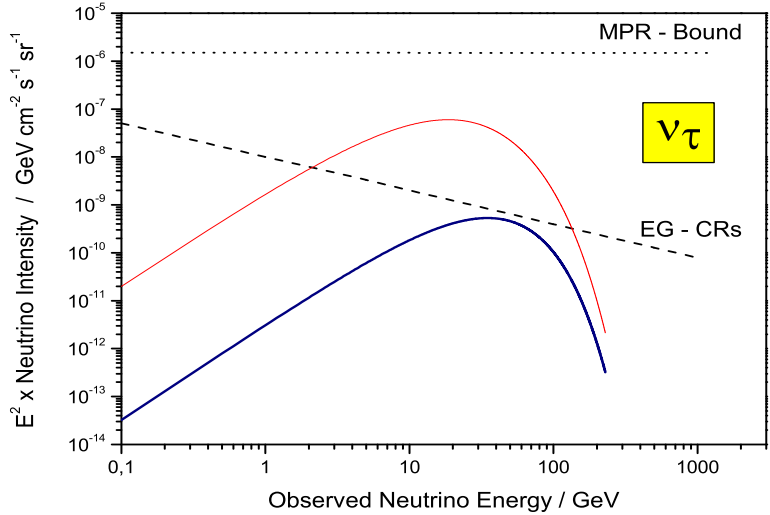}
	\end{center}
	\caption{Cosmological $\tau$-neutrino intensity over observed energy for a trial neutralino model and the Moore et al. profile; we plot the overall (upper solid line) and the estimated relatively local (up to z = 0.01; lower solid line) contribution in comparison. Also shown are the upper bound on the $\tau$-neutrino intensity from hadronic sources (dotted line) \cite{key-mpr,wm} and the estimated $\tau$-neutrino intensity due to cosmic ray interactions in external galaxies \cite{key-ml} (dashed line).}
	\label{fig:Fig4}
\end{figure}

\section{Neutrinos from neutralino annihilations}

The second very interesting channel is annihilation into high-energy neutrinos. The calculations presented
for the gamma rays in Section 2 also hold true for the neutrino
channels, taking into account some modifications: The absorption function
$\mathrm{\kappa}$ for neutrinos can be set equal to unity for the energy range
we are interested in. Also, the upper limit $\mathrm{z_{max}}$ for the integration
can be relaxed, and for our calculations was set to $\mathrm{z_{max}=10^{4}}$,
since for greater redshifts a $\mathrm{\mathcal{O}(100\,GeV)}$ neutrino is
redshifted into the overwhelmingly strong stellar background below
10 MeV. Branching ratios into neutrino producing channels show a wide range
of possible values. They can approach order unity
for $\mathrm{b\bar{b}}$-dominated channels, and can be of the order 0.1 for
$\mathrm{t\overline{t}}$ and also for $\mathrm{W^{+}W^{-}}$ processes. The differential neutrino
spectra for these processes also show some degree of variation.
Generally speaking, spectra will be "soft" (i.e. with a
shallow maximum at energies below one tenth of the neutralino mass)
for the $\mathrm{b\overline{b}}$-dominated channel, and harder, with a sudden
drop just below the $\mathrm{\chi^{0}_{1}}$-mass, for the $\mathrm{W^{+}W^{-}}$-process
\cite{key-25,key-nezri2}.

A robust approach to estimate the total neutrino yield from hadronization
and the following decay of charged $\pi$-mesons via $\mathrm{\pi\rightarrow\mu+\nu_{\mu}}$
and the following $\mathrm{\mu\rightarrow e+2\;{\nu}}\;$ 
is straightforward. For the total hadron spectrum from the annihilation,
one can use an approximation to the Hill spectrum \cite{key-27,key-28}, 
already taking into account that almost all produced hadrons are isospinsymmetric pions:\begin{equation}
\mathrm{\frac{dN_{\pi}}{d\eta_{\pi}}=\frac{10}{16}\eta_{\pi}^{-3/2}\left(1-\eta_{\pi}\right)^{2}}\label{eq:7}\end{equation}

with $\mathrm{\eta_{\pi}={E_{\pi}}/{m_{\chi}}}$. For the decay chains
mentioned above, each of these charged pions will produce three neutrinos / antineutrinos, and the neutrino multiplicity can be found by integration.  The branching ratios and differential neutrino spectra
can be obtained using \noun{DarkSusy}, \noun{SusSpect} \cite{suspect} and the \noun{Pythia/JetSet} Lund
Monte Carlo \cite{key-26} codes.
Due to neutrino oscillations, the $\,\nu_{e}\,:\,\nu_{\mu}\,:\,\nu_{\tau}\,$ ratio of the neutrinos arriving on earth will be \\{$\,1\,:\,1\,:\,1\,$ (max. mixing).
In Fig. 4, we plot the $\tau$-neutrino background for a trial 
$\mathrm{\chi\chi\rightarrow b\overline{b}}$ \cite{key-25}
- dominated model with
$\mathrm{\left\langle \sigma v\right\rangle =8.8\times10^{-27}\;{cm^{3}}\;{s}^{-1}}$ and a neutralino rest mass of 230 GeV. The Moore et al. case of halo profile is assumed. 
Rescaling of the local contribution to take the supergalactic plane into account is done according to the procedure for the gamma rays.
Again one can see that the foreground contributes a small but potentially
interesting part to the overall intensity. Since for electron and muon neutrinos, the atmospheric neutrino background will be approximately six orders of magnitude larger than our predicted intensity, we calculate the $\mathrm{\tau}$-neutrino intensity instead, for which the expected relevant backgrounds arise only from neutrino production in AGN sources and from cosmic ray interactions in external galaxies. It should be noted that the intensity from hadronic sources might well be weaker than the presented upper bound, making the annihilation induced intensity more significant. The cosmological neutrino background due to neutralino annihilations is therefore a new and interesting signature of the cold dark matter. In contrast to the well explored neutrino intensities arising from neutralino annihilations in the center of the Earth or the Sun, it is independent from the neutralino-nucleon scattering cross-section. If the neutralino is the WIMP, depending on the SUSY setup and strength of the AGN induced background, one would predict a characteristic "bump" in the extragalactic high-energy neutrino background at an energy corresponding to approximately ten percent of the neutralino mass. In future neutrino experiments, discrimination of $\tau$-neutrinos could be achieved by searching for a "double bang" signature (see e.g. \cite{db}) in the detector, albeit rather in the time structure of the PMT response rather than in the spatial distribution of light along the track.
\section{Discussion}

From the results presented above several conclusions can be drawn:

\begin{enumerate}
\item For the neutralino models we calculated, the gamma ray background due to WIMP annihilation can constitute a sizable fraction of the total extragalactic gamma ray intensity. The anisotropies in the WIMP-induced background are expected to be different from those in the astrophysical background (from faint and distant gamma ray emitting sources) due to the unique $\mathrm{\rho^{2}}$-dependence of the dark matter induced intensity.
\item Owing to the expansion of the Universe and structure evolution, for both
gamma rays and neutrinos a significant contribution to the overall
annihilation induced intensity is produced in the relatively local Universe. The $\gamma$-ray absorption at high redshifts sharpens this effect, in particular for neutralinos with TeV-scale masses.
\item Taking into account that the mass distribution in the local Universe
up to $\mathrm{z\approx0.01}$ is not isotropic but concentrated in
the so-called supergalactic plane \cite{key-29}, there should be
a large angular-scale intensity anisotropy on the level of some percent (depending on the SUSY scenario and halo profiles). Small scale anisotropies due to nearby galaxies and dark matter clumps in the Milky Way halo should also exist \cite{Baltz,olinto2}.
\item Using the halo profile and formalism given in \cite{Baltz}, one can compare the estimated gamma ray intensity produced in the dark matter halo of the cD-galaxy M87 (a possible target for indirect SUSY-searches) and the cosmological SUSY-induced intensity. We find that for a threshold energy of 50 GeV within a radius of 6 arc minutes from the center of M87 the cosmological intensity is about two orders of magnitude below the expected intensity from this individual halo. For targeted IACT observations, M87 should stand out very clearly from this irreducible background.
\item High energy neutrino intensities from cosmological neutralino annihilations
can be non-negligible when compared to other expected extragalactic neutrino intensities \cite{key-ml}, and should also show the discussed anisotropy. To make a detection of these WIMP-induced neutrinos feasible, one would need a flavor-discriminating $\nu_{\tau}$-detector (see e.g. \cite{kowalski,flavor}) to suppress the very strong atmospheric $\mathrm{\nu_{e}\;+\;\nu_{\mu}}$-background in the 10 GeV -- 1 TeV energy range. 
\end{enumerate}
To summarize, neutralino self-annihilation is a potentially interesting process to produce cosmological gamma ray and neutrino backgrounds. As a corollary, these irreducible backgrounds must not be neglected in indirect WIMP searches.
Further studies are demanding, in particular to incorporate mSUGRA-inspired regions of the parameter space into the scans and to rigorously asses the density perturbation evolution.

\begin{ack}
\sloppy Explicit thanks go to T. Kneiske for contributing data to include the gamma ray attenuation. D.E. especially wants to thank J. Edsjö and L. Bergström for enlightening discussions. Great thanks also go to J. Bösner for invaluable insights into numerical computing and to J. Niemeyer for proof-reading the manuscript.
Generous support by the BMBF Verbundforschung (O5CM0MG1) and the Helmholtz Gemeinschaft (VIHKOS) is gratefully acknowledged.
We thank the anonymous referee for helpful comments. 
\end{ack}

\end{document}